\documentclass[lettersize,journal]{IEEEtran}
\usepackage{amsmath,amsfonts}
\usepackage{algorithmic}
\usepackage{algorithm}
\usepackage{array}
\usepackage[caption=false,font=footnotesize,labelfont=rm,textfont=rm]{subfig}
\usepackage{textcomp}
\usepackage{stfloats}
\usepackage{url}
\usepackage{verbatim}
\usepackage{graphicx}
\usepackage{cite}
\usepackage{color}
\usepackage{algorithm}
\usepackage{float}

\usepackage{multirow}

\begin{document}

\title{
Can \textcolor{black}{Wireless Environmental Information} Decrease Pilot Overhead: A CSI Prediction Example
}
\author{Lianzheng Shi, Jianhua Zhang, Li Yu, Yuxiang Zhang, Zhen Zhang, Yichen Cai, and Guangyi Liu
\thanks{Lianzheng Shi, Jianhua Zhang, Li Yu, Yuxiang Zhang, and Yichen Cai are with the State Key Lab of Networking and Switching Technology, Beijing University of Posts and Telecommunications, Beijing 100876, China (e-mail: shilianzheng@bupt.edu.cn; jhzhang@bupt.edu.cn; li.yu@bupt.edu.cn; zhangyx@bupt.edu.cn; caiyichen@bupt.edu.cn).

Zhen Zhang is with the School of Electronic Information Engineering, Inner Mongolia University, Hohhot 010021, China(e-mail: zhenzhang@imu.edu.cn).

Guangyi Liu is with the Future Research Laboratory, China Mobile Research Institute, Beijing 100053, China (e-mail: liuguangyi@chinamobile.com).
}}

\markboth{Journal of \LaTeX\ Class Files,~Vol.~14, No.~8, August~2021}%
{Shell \MakeLowercase{\textit{et al.}}: A Sample Article Using IEEEtran.cls for IEEE Journals}


\maketitle

\begin{abstract}
\textcolor{black}{Channel state information (CSI) is crucial for massive multi-input multi-output (MIMO) system. 
As the antenna scale increases, acquiring CSI results in significantly higher system overhead.} 
\textcolor{black}{In this letter, we propose a novel channel prediction method which utilizes wireless environmental information with pilot pattern optimization for CSI prediction (WEI-CSIP).}
\textcolor{black}{Specifically, scatterers around the mobile station (MS) are abstracted from environmental information using multi-view images.} 
\textcolor{black}{Then, an environmental feature map is extracted by a convolutional neural network (CNN).} 
Additionally, the deep probabilistic subsampling (DPS) network acquires an optimal fixed pilot pattern.
\textcolor{black}{Finally, a CNN-based channel prediction network is designed to \textcolor{black}{predict} the complete CSI, using the environmental feature map and partial CSI.} 
\textcolor{black}{Simulation results show that the WEI-CSIP can reduce pilot overhead from 1/5 to 1/8, while improving prediction accuracy with normalized mean squared error reduced to 0.0113, an improvement of 83.2\% compared to traditional channel prediction methods.} 
\end{abstract}


\section{Introduction}
Antenna array scale is growing in the six-generation (6G) massive MIMO system, with the number of antennas increasing to the thousand-level, significantly improving spectrum efficiency\cite{EMIMO, zhang2023channel}. 
The \textcolor{black}{6G massive MIMO system} needs accurate CSI in this case. 
Besides, CSI is necessary for some communication tasks, such as user scheduling, beamforming, power allocation, and so on. 
\textcolor{black}{Conventional methods typically obtain CSI through channel estimation\cite{intelligent6G}. However, this approach results in a extremely high pilot overhead in 6G communication systems. It is challenging to acquire
highly accurate CSI with low overhead for the 6G system\cite{Zhang2023AI}.}

\textcolor{black}{Given that channel exhibit certain correlations across spatial, temporal, and frequency dimensions, leveraging AI methods to reduce overhead presents a promising approach\cite{GenerativeAI}.}
AI could dig out the deep relationships in space time frequency multi-domain channel which could reduce the necessary multi-dimensional channel source. 
In \cite{Oscar2024Comparison}, the authors \textcolor{black}{compares} many neutral network models adapted to channel prediction. 
A channel prediction method based on transformer is proposed and verified on the measured CSI data set in high-speed scenario~\cite{transformer_CSI}. 
In \cite{E2ENet}, the uplink pilot is leveraged directly to accomplish uplink channel estimation and downlink channel prediction. 
\textcolor{black}{In \cite{zhou2023lowoverhead}, the low dimensional feature vector is merged in MS, fed back to the base station (BS) and extrapolated to recover CSI to reduce the feedback overhead.} 
\textcolor{black}{In addition, the DataAI-6G channel dataset is continuously updated for channel prediction simulation validation\cite{data6G2023}.} 
\textcolor{black}{However, data-driven AI-enabled channel prediction suffers from its performance degradation in a complex environment, namely poor generalization of neural model. This leads to an increase in system overhead in order to meet system performance requirements. }

\textcolor{black}{Wireless signal propagation is influenced by the environment\cite{semantics_beam2023}. 
It is essential to set up a mapping relationship between the environment and the channel.} 
\textcolor{black}{With the rapid progress of sensing and AI technology, powerful tools are now available for acquiring and analyzing environmental information \cite{AIGC2024}.} 
Digital twin channel can construct the mapping relationship between the radio channel and the digital world, which is expected to significantly improve performance for 6G MIMO system \cite{wang2024DTC}.
Path loss is predicted by directly using the environmental information from the transmitter (TX), receiver (RX), and internal scatterers in the line-of-sight (LOS) case~\cite{sun2022environment_pathloss}.
\textcolor{black}{The environment semantics is defined as the spatial distribution of scatterers to achieve beam prediction by images in the MS\cite{Vision_Semantics}.} 
The radio environment knowledge pool illustrates the logical relationship between the physical world, the electromagnetic environment, and the wireless channel\cite{wang2023REKP}.
\textcolor{black}{Therefore, effectively incorporating environmental information into channel prediction can enhance performance by trading perceptual resources for communication resources, particularly in achieving low-overhead CSI prediction in MIMO system.} 

In this letter, we propose a novel channel prediction method termed wireless environmental information with pilot pattern optimization for CSI prediction (WEI-CSIP).
Specifically, each scenario can be provided with an optimal pilot pattern to facilitate the selection of partial pilots. By constructing the relationship between the environment and the channel, the image feature map can be obtained. Furthermore, the complete channel CSI is predicted using partial CSI and the environmental feature map. Since the WEI-CSIP uses special scenario optimization for each scenario and incorporates environmental information, it significantly improves the overall CSI prediction accuracy and lowers pilot overhead.

This paper is organized as follows. The environment and system model are introduced in Section II. In Section III, we detail our the WEI-CSIP and metrics. The simulation setup and results are given in Section IV. The conclusions are summarized in Section V.

\section{Environment and system model}
\subsection{Definition of environment scenario}
\textcolor{black}{
Consider an outdoor urban scenario with many cars on the road and buildings distributed randomly in several building groups. 
The variation of CSI is related to environmental information. 
To depict the surrounding environment as possible, multi-view images $I_1,I_2,…,I_{N_A}$ are captured by \textcolor{black}{several cameras} deployed at MS, where $I_{i}\in\mathbb{R}^{w\times h\times3}$ and $N_A$ is the number of camera viewpoints.
The panoramic image \textcolor{black}{$I\in\mathbb{R}^{(wN_{A})\times h\times3}$} is concatenated by these multi-view images, which can be expressed as:
\begin{equation}
I=[I_1,I_2,…,I_{N_A}]\text{.}
\label{four_image_stitch}
\end{equation}
}
Where $I_{1},I_2,…,I_{N_A}$ is sequentially the image data obtained at constant interval angle from the front respectively.

\subsection{System model}
\textcolor{black}{Consider orthogonal frequency division multiplexing (OFDM) modulation with \textcolor{black}{$N_c$} subcarriers and $N_T$ OFDM symbols in a frequency division duplexing (FDD) MIMO system \textcolor{black}{with a uniform line array (ULA) installed at the BS and MS. Assuming that there are $M_t$ transmitting antennas at the BS, and there are $N_r$ receiving antennas at the MS.} The received signal can be expressed as:}
\begin{equation}
\textcolor{black}{y_{m,n,p,q} = h_{m,n,p,q}x_{m,n,p,q}+n_{m,n,p,q}\text{.}}
\label{system_received_signal}
\end{equation}
Where $h_{m,n,p,q}$, $x_{m,n,p,q}$, and $n_{m,n,p,q}$ are the channel response, transmit signal, and Gaussian noise, respectively, of the \textcolor{black}{\emph{q}}th subcarrier of the \textcolor{black}{\emph{p}}th number of symbols of the \textcolor{black}{\emph{m}}th transmitting antenna and the \textcolor{black}{\emph{n}}th receiving antenna.

Let the channel response matrix in the time-frequency space of the \emph{m}th transmitting antenna and the \emph{n}th receiving antenna be \textcolor{black}{$\mathbf{H_{m,n}}\in{\mathbb{C}^{N_T\times{N_c}}}$}, which is denoted as:
\textcolor{black}{\begin{equation}
\mathbf{H_{m,n}}=\begin{bmatrix}h_{m,n,1,1}&h_{m,n,1,2}\quad\cdots&h_{m,n,1,N_{c}}\\\vdots&\vdots&\vdots\\h_{m,n,N_{T},1}&h_{m,n,N_{T},2}\ \cdots&h_{m,n,N_{T},N_{c}}\end{bmatrix}\text{.}
\label{full_H}
\end{equation}}



The goal is to predict the complete channel \textcolor{black}{$\mathbf{H}\in{\mathbb{C}^{N_T\times{N_c}\times{M_t}\times{N_r}}}$} using partial CSI and multi-view images.


\section{Environment-based Channel Pilot Pattern Optimization and Channel Prediction}
\subsection{The WEI-CSIP Architecture}
\begin{figure}[h]
\centerline{\includegraphics[scale=1]{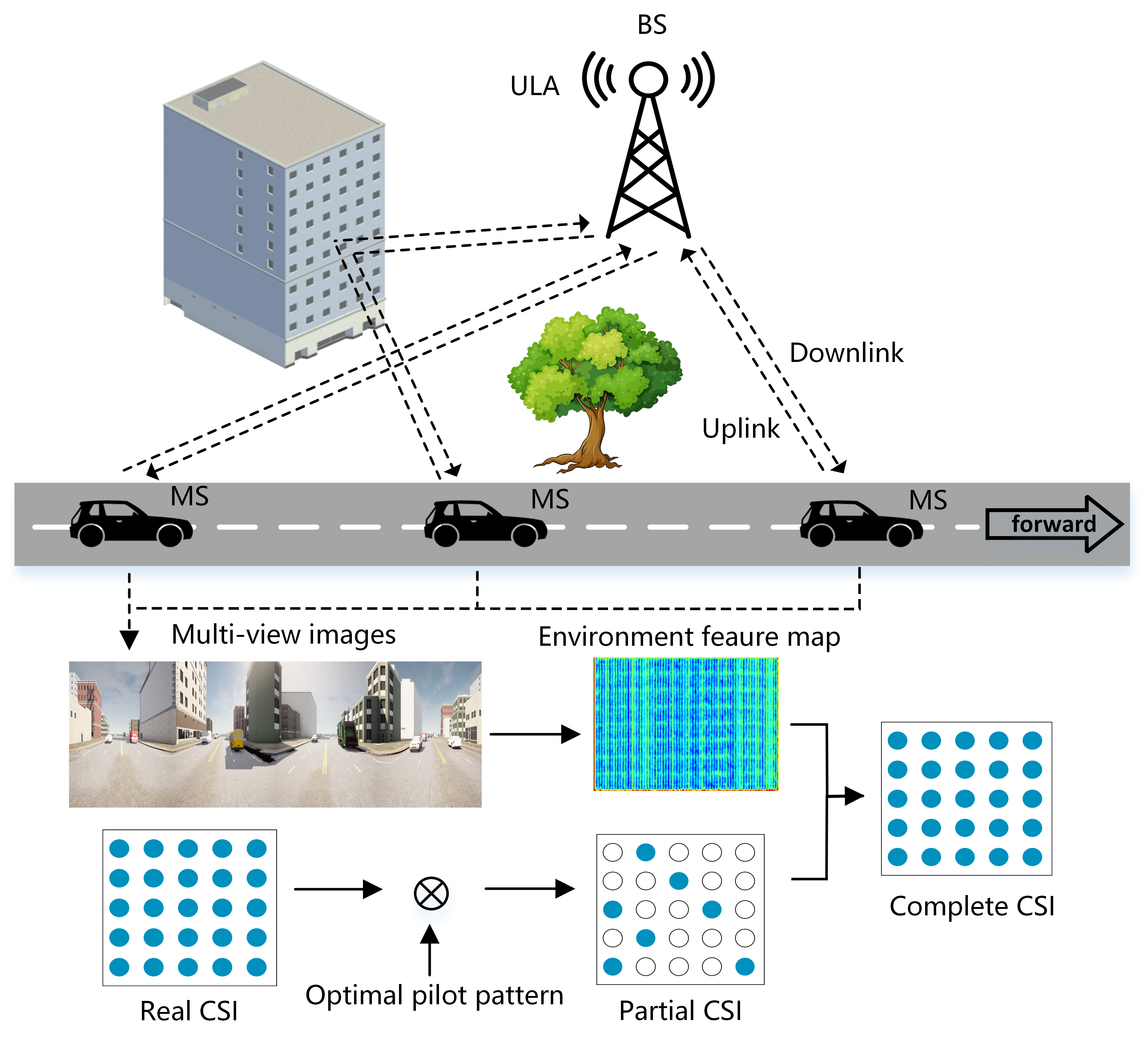}}
\caption{\textcolor{black}{Architecture of the WEI-CSIP.}}
\label{progress_structure_fig}
\vspace{-0.3cm}
\end{figure}
In order to accomplish channel prediction with high accuracy and low overhead, we propose a channel prediction method as shown in Fig. \ref{progress_structure_fig}. It is assumed that the \textcolor{black}{multi-view environmental images} can be obtained from the camera sensor.

\textcolor{black}{Environmental information is used as input for channel prediction, significantly reducing pilot overhead and predicting complete CSI at the BS.} The WEI-CSIP network consists of three modules including (1) the image feature extraction module, (2) the pilot pattern optimization module, and (3) the channel prediction module. 
\textcolor{black}{First, the image feature extraction module acquires the multi-view images and feeds them into a CNN for feature extraction. Simultaneously, the pilot pattern optimization module selects the optimal pilot pattern through the pilot pattern optimization network. Finally, based on the optimal pilot pattern and the image feature map, the channel prediction module reconstructs the complete channel matrix at the BS side.}
\textcolor{black}{For network model training, all modules need to participate in training to obtain the optimal pilot pattern for a specific scenario. Notably, the optimal pilot pattern is identical for each user in the scenario.}  

\subsection{Sub-neural Network Design}
In the following subsections, the three sub-network modules are described in detail.
\begin{figure*}[h]
\centerline{\includegraphics[scale=1.22]{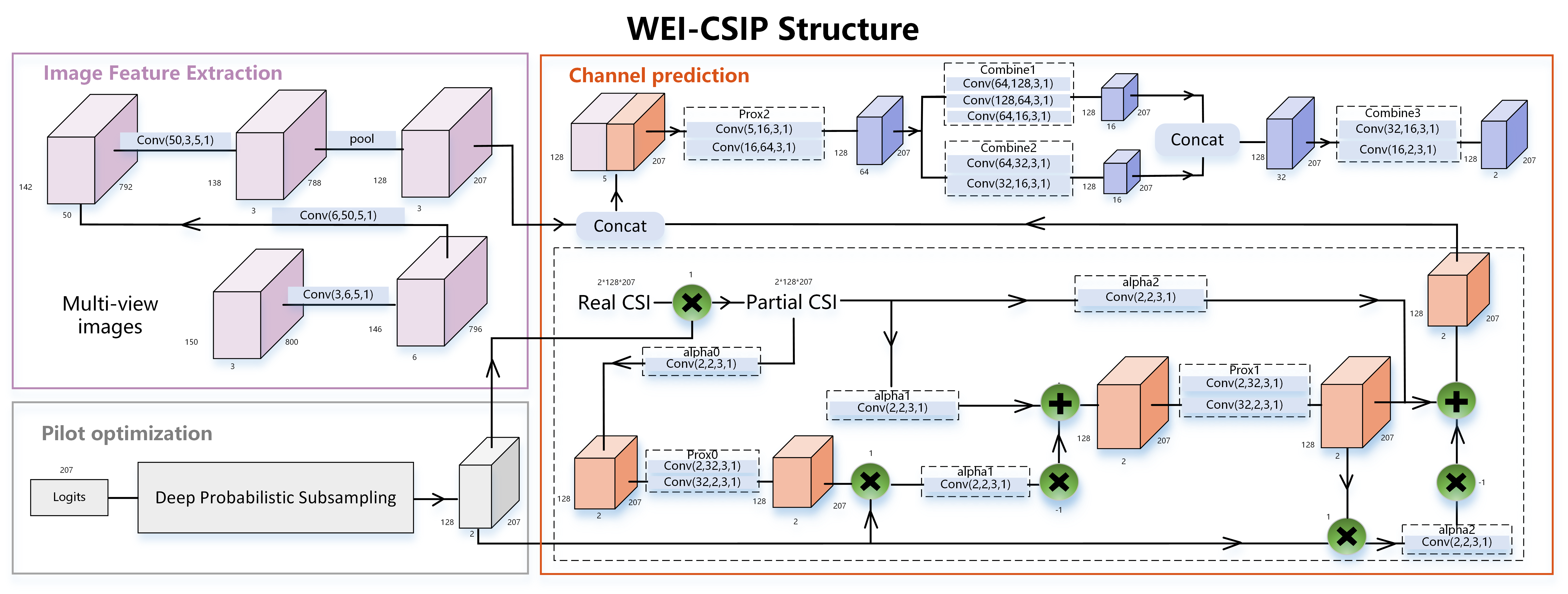}}
\caption{\textcolor{black}{Specific network design of the WEI-CSIP.}}
\label{model_structure_fig}
\vspace{-0.2cm}
\end{figure*}

\subsubsection{Image feature extraction module}
\textcolor{black}{CNN excels in visual tasks due to its ability to extract hierarchical features from images. 
Multi-view images can depict the surrounding scenario environmental information of the receiver.}
As shown in Fig. \ref{model_structure_fig}, three convolutional layers and one pooling layer are used to extract the image feature map from panoramic image $I$. The image feature map represents the potential mapping relationship between the local environmental information and CSI.

\subsubsection{Pilot pattern optimization module}
\textcolor{black}{Low-overhead optimal pilot pattern design can be formulated as a task-adaptive compressed sensing problem, where the goal is to select the optimal subset of signal samples to enable end-to-end optimization. }
\textcolor{black}{For each $\mathbf{H_{m,n}}$ corresponding to the \emph{m}th transmitting antenna and \emph{n}th receiving antenna, the optimal pilot pattern $\mathbf{A}\in{\{0,1\}^{{N_T\times{N_c}}}}$ in the time-frequency dimension is generated by the DPS \cite{kool2019stochastic}. }
\textcolor{black}{$\mathbf{H}_{partial}\in{\mathbb{C}^{N_T\times{N_c}}}$ is the CSI at a given pilot pattern in the time-frequency dimension, which can be expressed as follows:}

\begin{equation}
\textcolor{black}{\mathbf{H}_{partial}={\mathbf{H_{m,n}}}\mathbf{A}\text{.}}
\label{H_partial}
\end{equation}
Where the number of elements with 1 in the optimal sampling matrix $\mathbf{A}$ is equal to the number of pilots $N_p$. The optimal pilot pattern is acquired during forward propagation as shown in Fig. \ref{model_structure_fig}. \textcolor{black}{The $softmax_\tau$(·) function used as the soft sampling matrix calculates the gradient during network model training in backpropagation.} 
\textcolor{black}{The pilot pattern optimization module provide optimal pilot pattern for channel prediction module to achieve higher prediction performance.} 
\subsubsection{Channel prediction module}
\textcolor{black}{The channel prediction module takes the image feature map and channel matrix $\mathbf{H}_{partial}$ as inputs. It predicts the complete channel matrix $\hat{\mathbf{H}}$ using a neural network combined with a proximal iterative algorithm \cite{parikh2014proximal}, which accelerates model convergence.} The iterative steps are denoted as:
\textcolor{black}{\begin{equation}
\begin{aligned}
&\mathbf{S^{k+1}}&&:=\quad\begin{cases}\mathbf{\alpha^0}{\mathbf{H}_{partial}}(k=0)\,;\\\mathbf{X^k}+\alpha^k(\mathbf{H}_{partial}-\mathbf{X^k}*\mathbf{A})(\mathrm{k>0})\end{cases}\\&\mathbf{X^{k+1}}&&:=\quad\mathbf{prox}_k(\mathbf{S^{k+1}})\,.
\label{Proximal_Algorithm}
\end{aligned}
\end{equation}}\textcolor{black}{Where $\mathbf{X^k}$, $\mathbf{prox_k}$, $\alpha^k$ are the forward value, proximal operator, and extrapolation parameters for the (${k+1}$)th iteration, respectively. $\mathbf{X^{k+1}}$ and $\mathbf{S^{k+1}}$ are the backward value and intermediate variables of the (${k+1}$)th iteration, respectively. }
We use a convolutional layer as the proximal operator $\mathbf{prox_k}$ and the prediction parameter $\alpha^k$. 
\textcolor{black}{The number of iterations of the algorithm is set to $N_o$. The CSI data obtained after iteration is combined with the number of channels with the image feature map. The complete CSI prediction is obtained by the well-designed CNN network. The specific network design is shown in Fig. \ref{model_structure_fig}.}

The $\hat{\mathbf{H}}$ predicted by the WEI-CSIP network can be expressed as:
\begin{equation}
\hat{\mathbf{H}}={\rm F}(\mathbf{H}_{partial},I)\text{.}
\label{H_partial}
\end{equation}
Where F is the model function of the WEI-CSIP network, and the mean square error (MSE) of the predicted and true values is regarded as the network loss function, i.e:
\begin{equation}
\textcolor{black}{
L_H=\frac1{{M_{t}}{N_{r}}}\sum_{i=1}^{M_{t}}\sum_{j=1}^{N_{r}}||\mathbf{H}_{i,j}-\hat{\mathbf{H}}_{i,j}||^2\text{.}
}
\label{MSE}
\end{equation}
Normalized mean squared error (NMSE) is a commonly used performance metric to assess the degree of error between the estimation or reconstruction results and the true value. For the predicted $\hat{\mathbf{H}}$ and the true value $\mathbf{H}$, the NMSE is defined as follows:
\begin{equation}
\textcolor{black}{NMSE=\frac{\sum_{i=1}^{M_{t}}\sum_{j=1}^{N_{r}}||\mathbf{H}_{i,j}-\hat{\mathbf{H}}_{i,j}||^2}{\sum_{i=1}^{M_{t}}\sum_{j=1}^{N_{r}}||\mathbf{H}_{i,j}||^2}\text{.}
\label{NMSE}}
\end{equation}
\textcolor{black}{Moreover, the model performance is also evaluated by calculating the cosine similarity between the true CSI and the predicted CSI. It is defined as:}
\begin{equation}
\textcolor{black}{Cosine\;Similarity=\frac{\sum_{i=1}^{M_{t}}\sum_{j=1}^{N_{r}}(\mathbf{H_{i,j}}\cdot\hat{\mathbf{H}}^*_{i,j})}{\sum_{i=1}^{M_{t}}\sum_{j=1}^{N_{r}}\|\mathbf{H}_{i,j}\|\|\hat{\mathbf{H}}_{i,j}\|}\text{.}}
\label{Cosine Similarity}
\end{equation}

\section{Simulation results}
\subsection{Simulation setup}
\begin{figure}[htbp]%
    \centering
    \subfloat[The simulation scenario for environment image.]{
        \label{origin_scenario_carla}
        \includegraphics[width=0.65\linewidth]{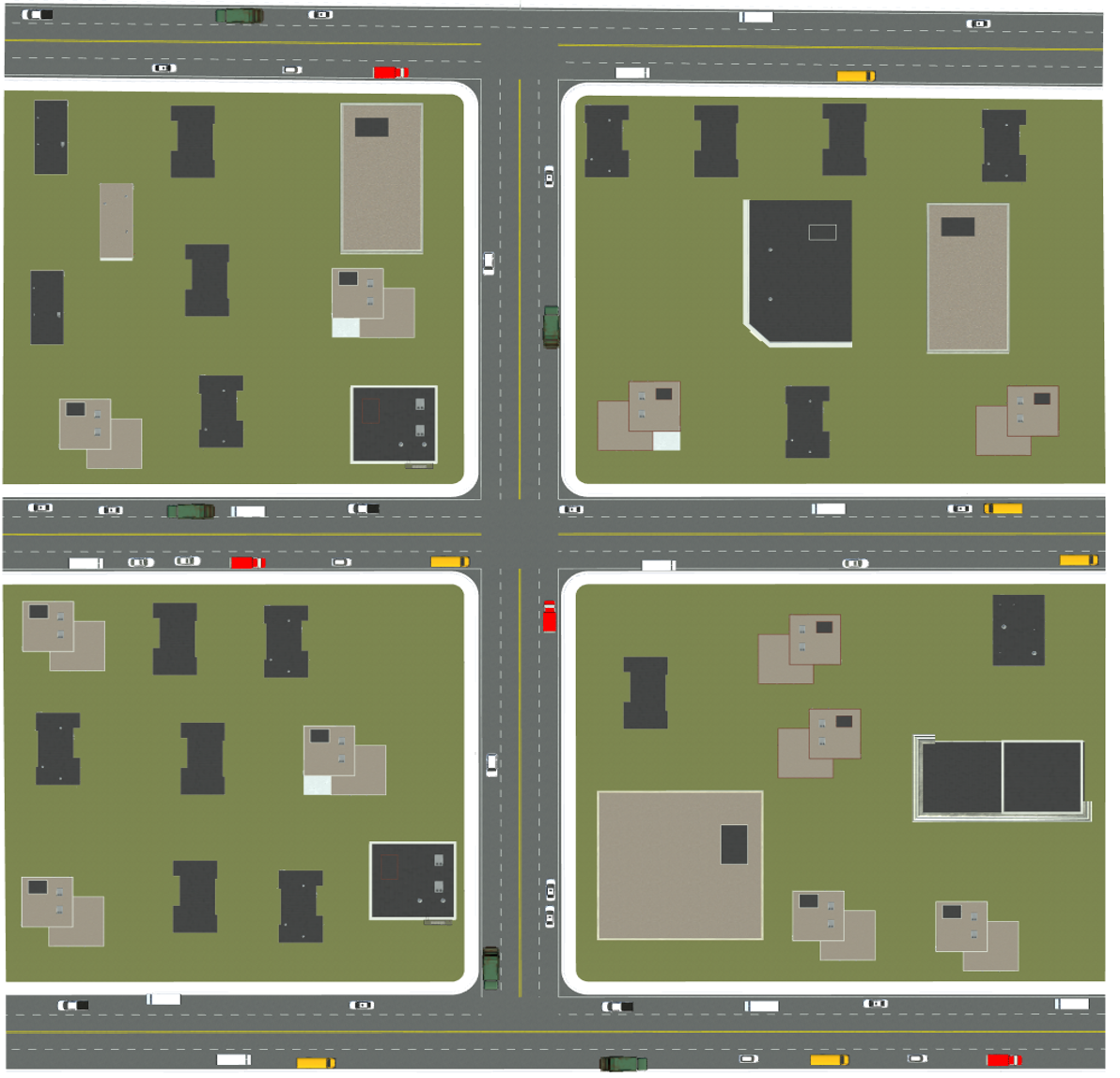}
        }\hfill
    \subfloat[The ray-tracing simulation scenario.]{
        \label{origin_scenario_wi}
        \includegraphics[width=0.9\linewidth]{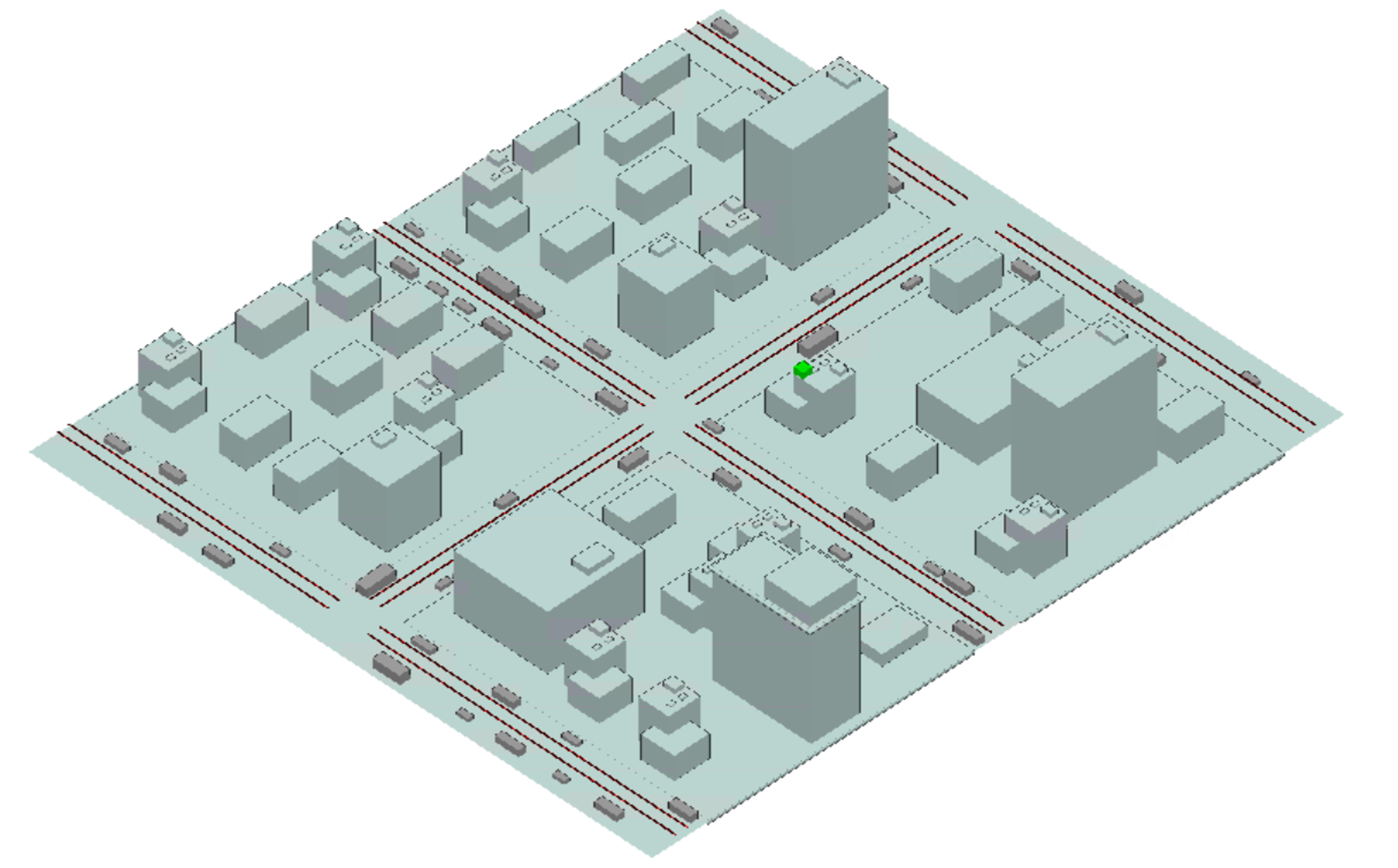}
        }
    \caption{\textcolor{black}{The origin environment and channel simulation scenarios.}}
    \label{origin_scenario}
\end{figure}

As shown in Fig.\ref{origin_scenario}, we construct a large outdoor urban scenario with a length of $L$ and a width of $W$ in the autonomous driving simulation software CARLA\cite{dosovitskiy2017carla} and set a BS to cover the \textcolor{black}{scenario}. The scene contains four \textcolor{black}{building groups} and \textcolor{black}{four roads}, and different types of vehicles are randomly placed on the road located on one side of the road to ensure that the scene has enough diversity. Vehicles moving on the road are simulated by deploying cameras at intervals of $L_c$ on the \textcolor{black}{roads} and sequentially capturing multi-view images. As the loss of surface detail has a limited effect on the channel, the entire scene is imported into blender software for simplification, where buildings and vehicles are replaced with simple cubes. The simplified model of the scene is imported into Wireless insite for ray-tracing simulation. Through the above steps, a dataset of multi-view images and CSI is constructed at different locations in this urban scene.

We set $L$=200 m, $W$=200 m, and $L_c$=0.26 m; the ULA of the BS is fixed 1m above the house in the center of the scene, with a height of 19m; the single antenna of the MS is set in \textcolor{black}{the roads}, with a spacing of $L_c$ of 0.26 m; the cameras are installed horizontally with $N_A$=4, and $w$×$h$=150×200 in pixels. The \textcolor{black}{ray-tracing} simulation parameter settings are shown in Table \ref{RT_parameter}; $M_{t}$=128, \textcolor{black}{$N_{r}$=1}, $N_c$=69, $N_T$=3, $N_{p}=\frac{N_cN_T}{8}$=26. Then we divide the dataset into a training set, a test set, and a validation set by 7:1:2, in which the data of the \textcolor{black}{four roads} are taken in turn and randomly disordered. The training set contains 3931 samples, the test set contains 562 samples, and the validation set contains 1123.
\textcolor{black}{The learning rate is set to 0.0008 for the image feature extraction network, 0.008 for DPS network, and 0.0008 for the channel prediction network. The coefficient for the LeakyReLU layer is set to 0.0005, the temperature coefficient for the DPS network is set to 2, and the $\rm{N_o}$ is set to 2. }
\begin{table}[htbp]
\vspace{-2.0em}
\setlength{\tabcolsep}{11pt} 
\renewcommand\arraystretch{1.2}  
\caption{ray tracing parameters for wireless insite.}
\label{RT_parameter}
\begin{center}
\fontsize{9}{9}\selectfont 
\begin{tabular}{|m{3.5cm}<{\centering}|m{3.5cm}<{\centering}|}
\hline
 {Middle frequency} & {6775 MHz} \\
\hline
 {Bandwidth} & {100 Mhz} \\ 
\hline
 {Subcarrier} & {120 kHz}  \\
\hline
 {Pilot selected proportion} & {1/12}  \\
\hline
 {OFDM symbol number} & {3}  \\
\hline
 {\textcolor{black}{Antenna number}} & {128}  \\
\hline
 {Reflection order} & {6}  \\
\hline
 {Diffraction order} & {1}  \\
\hline
 {Path amount reserved for each Rx} & {\textcolor{black}{15}}  \\
\hline
 {Tx location} & {In center, 1 m above the roof
height: 19 m}  \\
\hline
 {Rx location} & {\centering height: 2 m \par \centering gap: 0.26 m \par Traversal in the street}  \\
\hline
\end{tabular}
\end{center}
\vspace{-2.0em}
\end{table}
\subsection{Numerical results}
Based on the optimal hyperparameter settings, we evaluate the prediction performance of the proposed the WEI-CSIP model by comparing the WEI-CSIP with the other three methods. 
\textcolor{black}{Namely random sampling without environmental information (RSWOEI), random sampling with environment (RSWEI), and DPS without environmental information (DWOEI). The RSWOEI and DWOEI only use \textcolor{black}{partial CSI} to predict complete CSI. The RSWEI incorporates multi-view images to extract the environmental feature map to predict complete CSI.} 
We use the RSWOEI as a baseline. 
For fairness, all of them are trained for \textcolor{black}{200} epochs, using the same loss function $L_\mathbf{H}$ in \ref{MSE}. 
NMSE and cosine similarity are used to assess the accuracy and reliability of the channel prediction model. 
\begin{figure}[htbp]%
    \centering
    \subfloat[\textcolor{black}{The NMSE of the WEI-CSIP and baseline.}]{
        \label{NMSE}
        \includegraphics[width=1\linewidth]{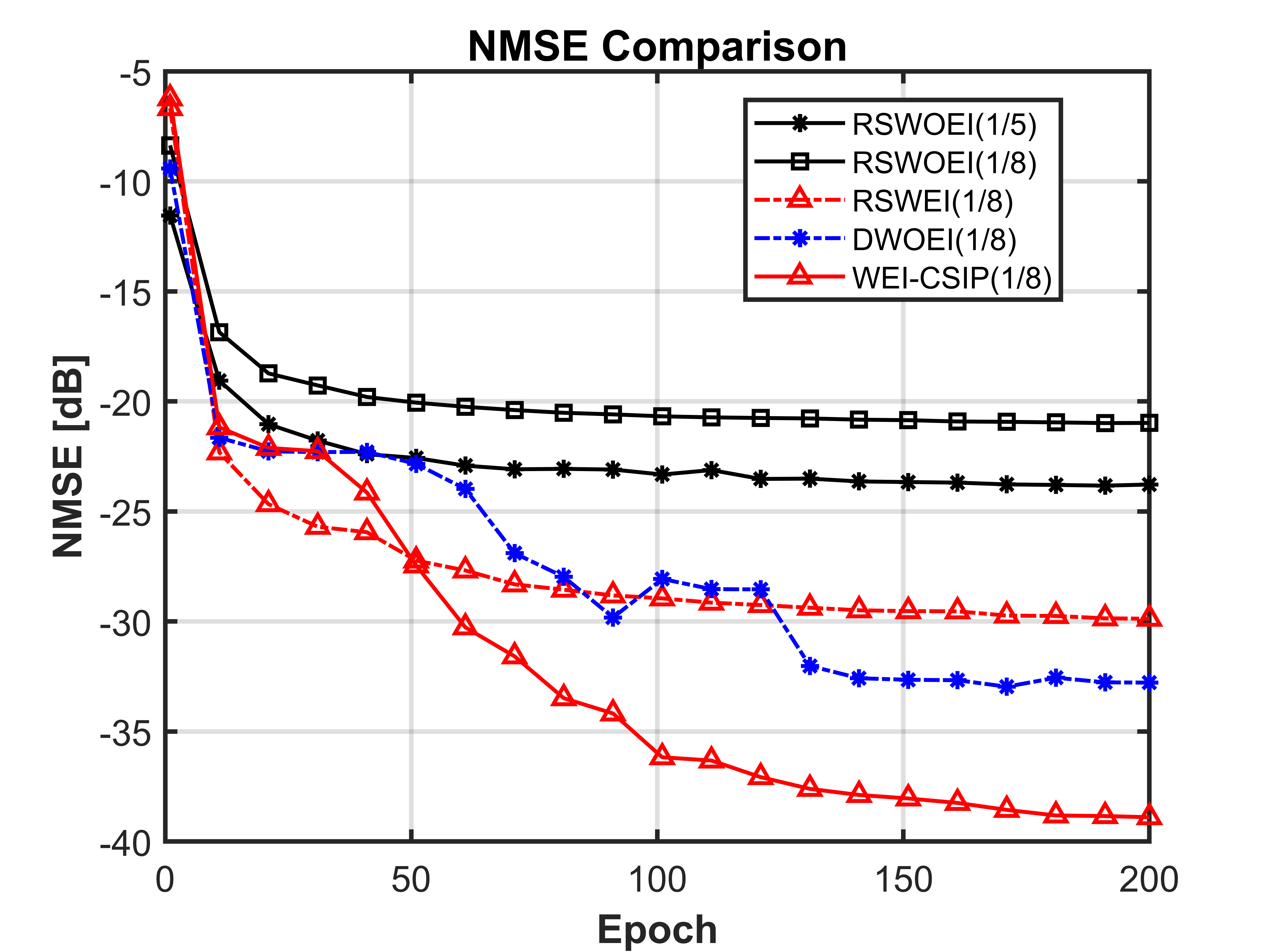}
        }\hfill
    \subfloat[\textcolor{black}{The cosine similarity of the WEI-CSIP and baseline.}]{
        \label{Cosine Similarity}
        \includegraphics[width=1\linewidth]{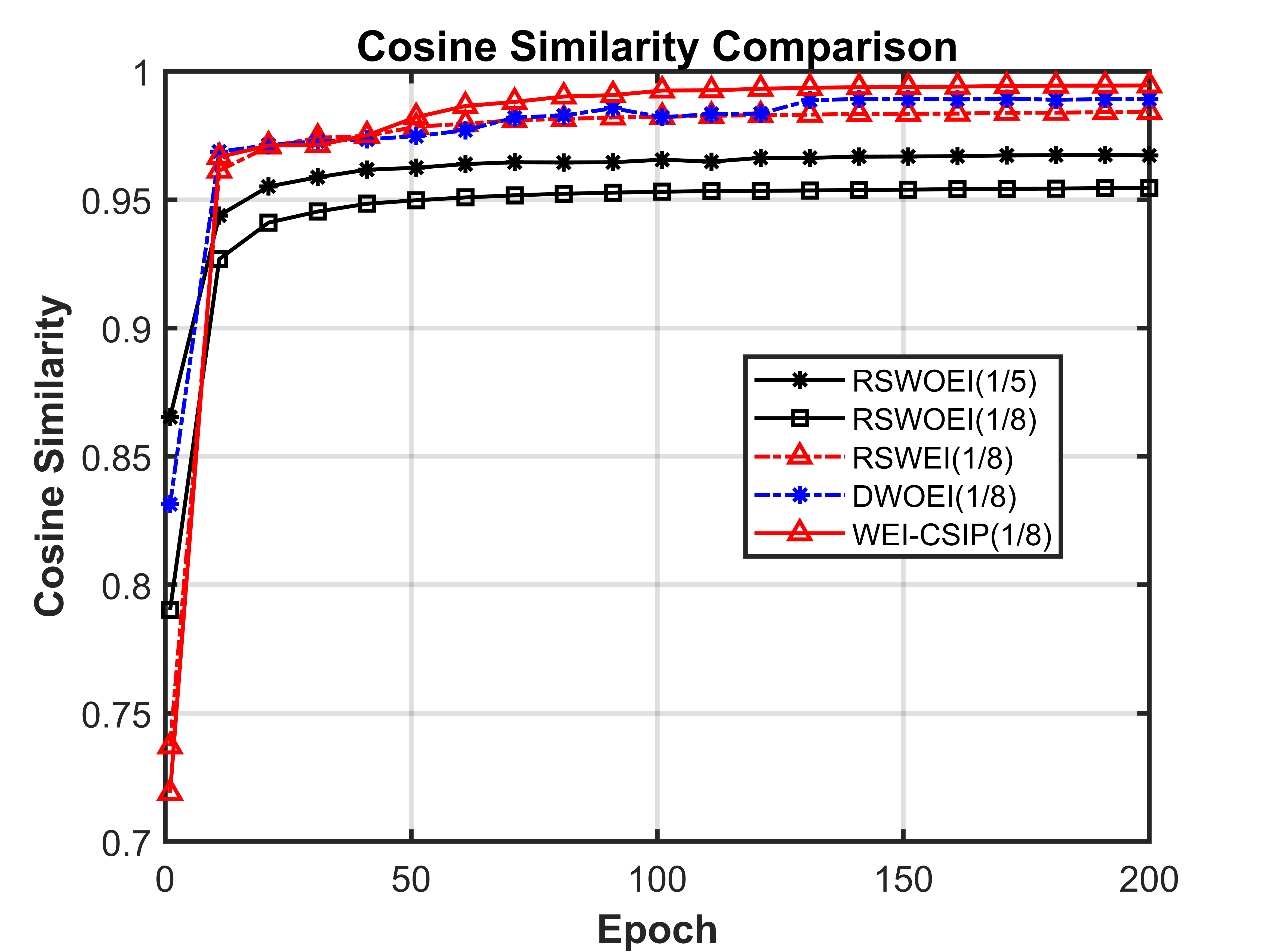}
        }
    \caption{\textcolor{black}{The performance of the WEI-CSIP and baseline.}}
    \label{performance_curve}
\end{figure}


\textcolor{black}{As shown in Fig. (\ref{NMSE}) and Fig. (\ref{Cosine Similarity}), the WEI-CSIP achieves the highest correlation coefficient and the lowest NMSE during model training of the four methods. }
\textcolor{black}{The results are analyzed below by selecting the number of 1/8 pilots. }
\textcolor{black}{Comparing the results of WEI-CSIP with RSWEI, or DWOEI with RSWOEI, the optimal pilot pattern is beneficial to improve CSI prediction accuracy in the same environment, leading to reducing NMSE by about 64.7\% and 76.4\%. In addition, Comparing the results of WEI-CSIP with DWOEI, or RSWEI with RSWOEI, it can be seen that the environmental information brings gain, resulting in a reduction of the NMSE by about 46.4\% and 64.2\%.} \textcolor{black}{Furthermore, compared with RSWOEI at the number of 1/5 pilots, the performance of CSI prediction is better with the number of 1/8 pilots by WEI-CSIP. Therefore, the pilot overhead is reduced by at least 37.5\%.}

\begin{figure}[!ht]
\centerline{\includegraphics[scale=0.3]{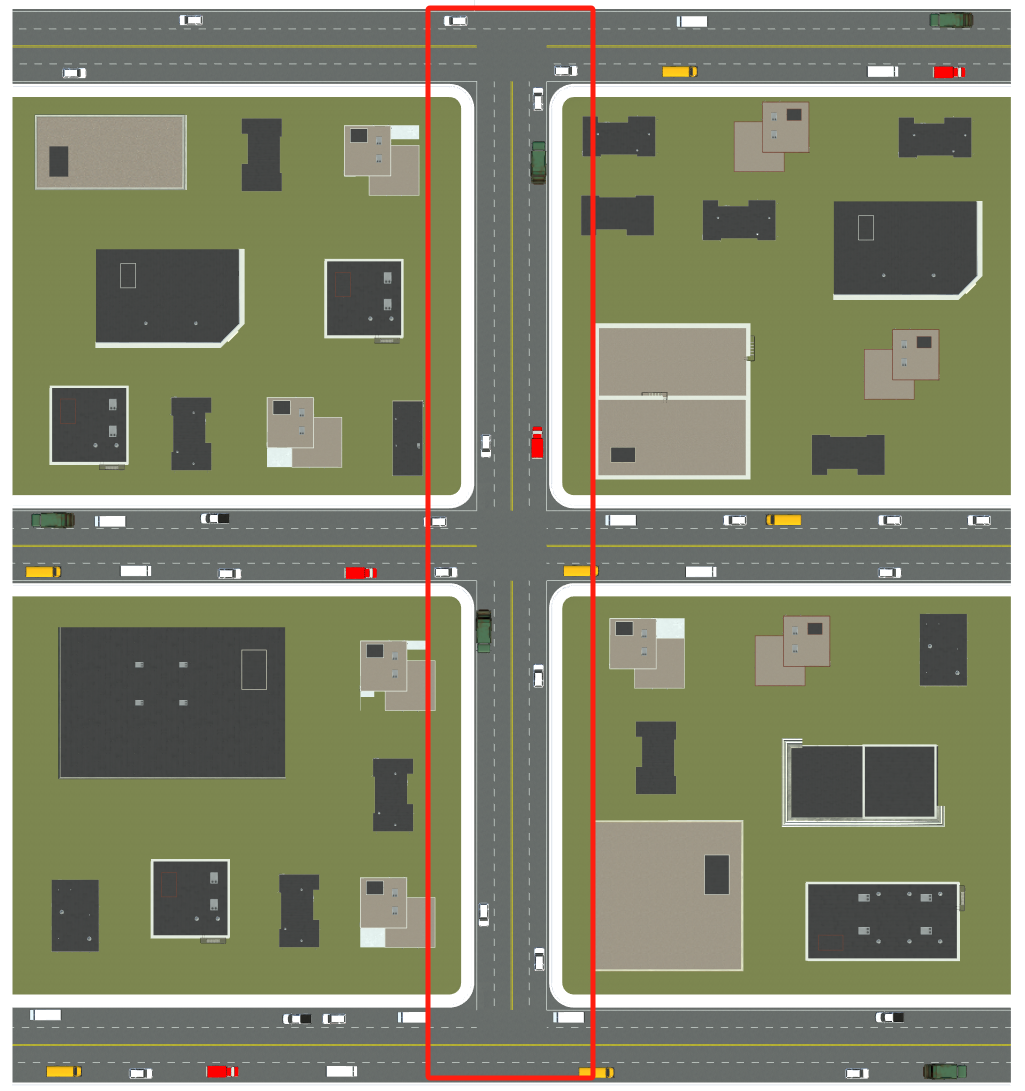}}
\caption{New scenario used to validate generalization. \textcolor{black}{All Rx are set in the red box range.}}
\label{new_sense}
\end{figure}
In addition, we modify the building and vehicle layout of the original scene to create a new scene, as shown in Fig \ref{new_sense}. The new dataset contains 1601 Rx for the middle two roads \textcolor{black}{(red box)}. The trained optimal models from four methods are tested on the new dataset. The results are summarised in Table \ref{new_sense_comparison}. Overall, the WEI-CSIP cosine similarity and NMSE in the new scenario are similar to the original scenario, indicating that the network model generalizes well. 

\begin{table}[!t]
\setlength{\tabcolsep}{11pt} 
\renewcommand\arraystretch{1.2}  
\caption{\textcolor{black}{The model performance of comparison in new and origin sense.}}
\begin{center}
\fontsize{9}{9}\selectfont 
\begin{tabular}{|m{2cm}<{\centering}|m{0.7cm}<{\centering}|m{0.7cm}<{\centering}|m{0.7cm}<{\centering}|m{0.7cm}<{\centering}|}
\hline
 \multirow{2}{*}{Methods} & \multicolumn{2}{c|}{\textbf{cosine similarity}} & \multicolumn{2}{c|}{\textbf{NMSE}}\\
\cline{2-5}
 & {origin}  & {new} & {origin} & {new}\\
\hline
 {RSWOEI} & {0.9545}  & {0.9478} & {0.0893} & {0.1030}\\
\hline
 {RSWEI} & {0.9841}  & {0.9796} & {0.0320} & {0.0414}\\
\hline
 {DWOEI} & {0.9899}  & {0.9840} & {0.0211} & {0.0328}\\
\hline
 {WEI-CSIP} & {0.9945}  & {0.9917} & {0.0113} & {0.0171}\\
\hline
\end{tabular}
\end{center}
\label{new_sense_comparison}
\vspace{-2.0em}
\end{table}

\section{conclusion}
\textcolor{black}{In this paper, we propose a novel channel prediction method designed for the MIMO system that reduces pilot overhead.} 
\textcolor{black}{Leveraging environmental information for the first time and partial CSI in the time-frequency dimension, the WEI-CSIP predicts the complete CSI.} 
Compared with the existing channel prediction methods, the WEI-CSIP incorporates the environmental information and optimal pilot pattern. 
\textcolor{black}{Simulation results show that the NMSE gain from environmental information can be up to 64.2\%, and the NMSE gain from the DPS can be up to 76.4\%. When reducing the pilot overhead by 37.5\%, the WEI-CSIP improves prediction accuracy by 82.3\% compared to RSWOEI.}
Moreover, the WEI-CSIP can achieve stable accuracy in different scenarios, declining the pilot overhead. 
\textcolor{black}{The WEI-CSIP achieves pleasant performance with lower overhead, reflecting the potential prospect of incorporating environmental information into channel prediction. } 
\section{ACKNOWLEDGMENT}

\bibliographystyle{IEEEtran}
\bibliography{cite}
\end{document}